\documentclass[twoside,a4paper,11pt]{torus2012}
\usepackage{graphicx}
\def\brfrac#1#2{\left(\dfrac{#1}{#2}\right)}
\def\dfrac#1#2{{\displaystyle\frac{\mathstrut #1}{#2}}}

\begin{document}
\pagenumbering{arabic}
\pagestyle{myheadings}
\thispagestyle{empty}
{\flushleft\includegraphics[width=\textwidth,bb=58 650 590 680]{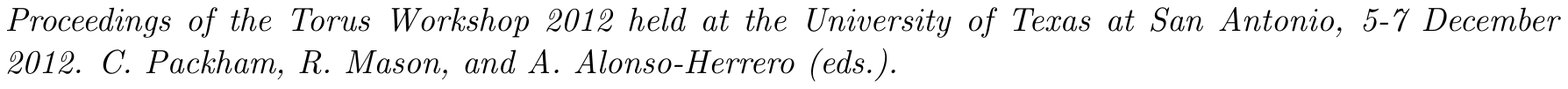}}
\vspace*{0.2cm}
\begin{flushleft}
{\bf {\LARGE
%
Innermost structure and near-infrared radiation 
of dusty clumpy tori in active galactic nuclei
%
}\\
\vspace*{1cm}
%
Toshihiro Kawaguchi
%
}\\
\vspace*{0.5cm}
%
Dept.\ of Physics and Information Science, Yamaguchi University, 
Yoshida 1677-1, 
Yamaguchi 753-8512, Japan
%
\end{flushleft}
%
\markboth{
Innermost structure and NIR reverberation of AGN dusty clumpy tori
%
}{ 
%
Toshihiro Kawaguchi
%
}
\thispagestyle{empty}
\vspace*{0.4cm}
\begin{minipage}[l]{0.09\textwidth}
\ 
\end{minipage}
\begin{minipage}[r]{0.9\textwidth}
\vspace{1cm}
\section*{Abstract}{\small
%
The dusty clumpy torus surrounds the central black hole 
(BH) and the accretion disk in active galactic nuclei, and 
governs the growth of super-massive BHs via gas fueling 
towards the central engine. Near-infrared (NIR) monitoring 
observations have revealed that the torus inner radius is 
determined by the dust sublimation process. However, the 
observed radii are systematically smaller than the theoretical 
predictions by a factor of three.

  We take into account the anisotropic illumination by the 
central accretion disk to the torus, and calculate the 
innermost structure of the torus and the NIR time variablity.
We then show that the anisotropy naturally solves the 
systematic descrepancy and that the viewing angle is the 
primary source to produce an object-to-object scatter of 
the NIR time delay. Dynamics of clumps at the innermost 
region of the torus will be unveiled via future high-resolution 
X-ray spectroscopy (e.g., Astro-H).
%
\normalsize}
\end{minipage}
%
%
%
\section{Introduction \label{intro}}
Active galactic nuclei (AGNs) are powered by gas accretion onto
supermassive black holes (BHs) at the center of each galaxy.
A variety of observations suggest that
the accretion disk and the BH are surrounded by an optically and
geometrically thick torus
 (e.g., \cite{Telesco84}, \cite{Antonucci85}). 
Since the 
torus potentially plays a role of a gas reservoir 
for the accretion disk, 
its nature, 
such as the structure, 
the size and the mass, 
has long been investigated 
(\cite{Pier92}, \cite{Pier93},\cite{Fukue93},  
\cite{Granato94}, 
\cite{Efstathiou95}.

A large geometrical thickness of the torus 
revealed by various observations
(\cite{Antonucci93}, 
 \cite{Pogge89}, 
 \cite{Wilson94}, \cite{Schmitt96}) 
 indicates
that numerous dusty clumps, 
rather than a smooth mixture of gas and dust, 
constitute the torus with a large clump-to-clump velocity 
dispersion 
(\cite{Krolik88}). 
Temperature of clumps is less than a critical 
temperature $T_{\rm sub} \sim 1500$\,K above which 
dust grains 
are sublimated (\cite{Barvainis87}).

Based on the energy balance of the clump closest to the BH, 
Barvainis (\cite{Barvainis87}) derived the innermost radius of the torus
(dust sublimation radius, denoted as $R_{\rm sub,0}$ in this
study):  
\begin{equation}
R_{\rm sub,0} = 0.13 \brfrac{L_{\rm UV}}{10^{44} \, \mathrm{erg}\,\mathrm{s}^{-1}}^{0.5}
 \brfrac{T_{\rm sub}}{1500 \,\mathrm{K}}^{-2.8}
 \brfrac{a}{0.05 \,\mu \mathrm{m}}^{-0.5} {\rm pc},
 \label{eq:bar87}
\end{equation}
where $L_{\rm UV}$ 
and $a$ are 
UV luminosity 
and the size of dust grains, respectively.

\begin{figure}
\center
\includegraphics[scale=0.55,bb=59 198 771 560]{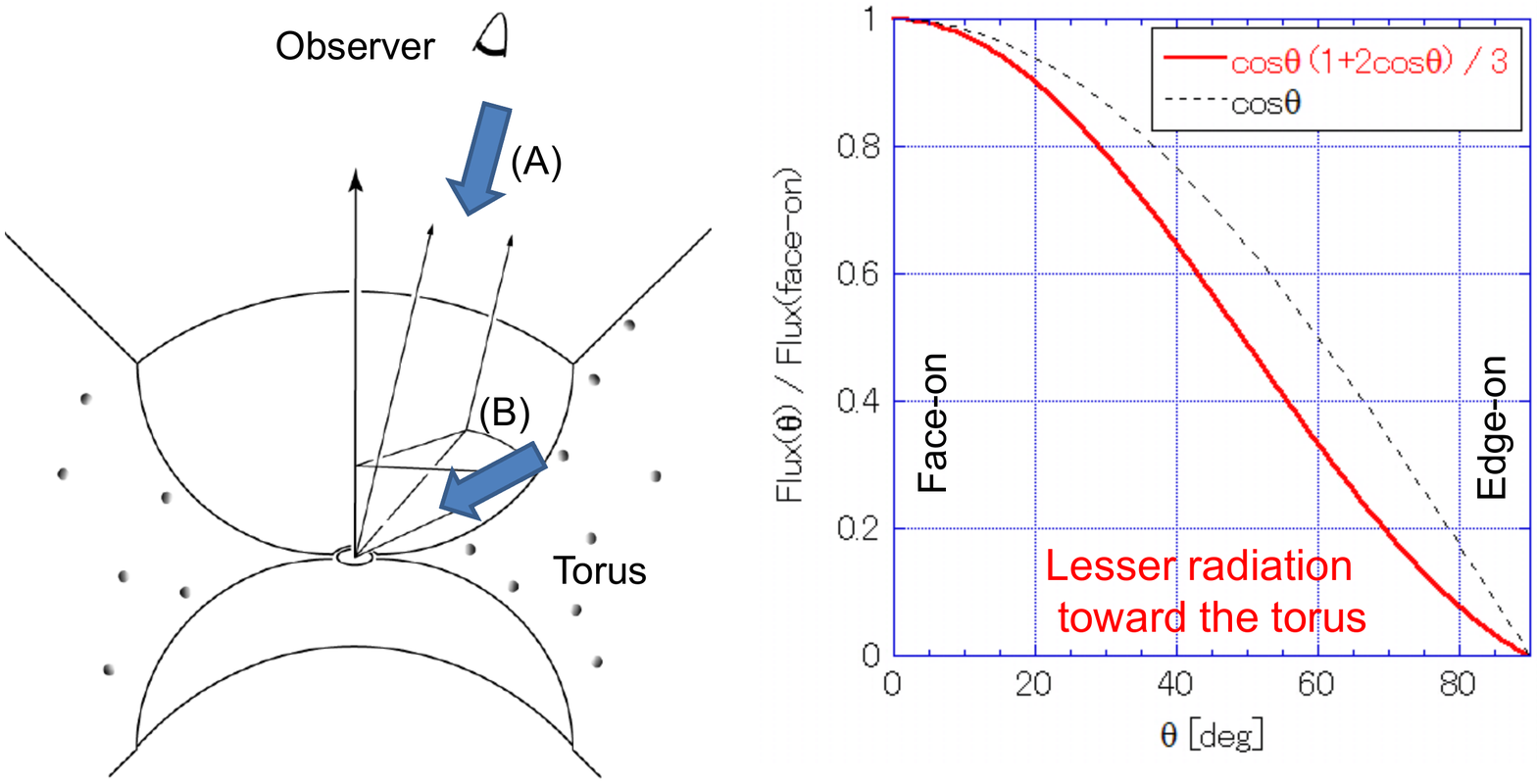}
\caption{\label{fig1} 
Left: (A) The inclination angle 
at which we observe the disk in type-1 AGNs is 
systematically different from (B) the angle 
at which an aligned torus observes the disk \cite{Kawaguchi10}. 
Right: Flux changes with the direction \cite{Kawaguchi11}. 
We must be aware that the flux towards the torus (B) 
is much weaker than that to the observer (A).
}
\end{figure}

Indeed, Near-infrared (NIR) emission 
from type-1 AGNs lags behind optical variation by an order of 
$R_{\rm sub,0} / c$
(\cite{Clavel89}, \cite{Glass92}, \cite{Glass04}, 
\cite{Nelson96}, \cite{Oknyanskij99},  
\cite{Minezaki04}, \cite{Suganuma04}).
Moreover, the luminosity dependency of the time lag also 
coincides with the theoretical prediction as $\propto L_{\rm UV}^{0.5}$
(\cite{Suganuma06}). 
However, the NIR-to-optical time lag 
is systematically smaller than the lag predicted from 
Equation (\ref{eq:bar87}) by a factor of $\sim \! 1/3$ 
(\cite{Oknyanskij01}, \cite{Kishimoto07}, \cite{Nenkova08}).
To tackle with this conflict, Kawaguchi \& Mori (\cite{Kawaguchi10}) 
pointed out that the illumination by 
an optically thick disk is inevitably anisotropic, 
which is a fact missing in deriving Equation (\ref{eq:bar87}).
There is a systematic difference between (A) the inclination angle 
at which we observe the disk in type-1 AGNs and (B) the angle 
at which an aligned torus observes the disk (Figure~1, left).
The effects of the anisotropic 
illumination 
naturally resolve
the puzzle of the systematic deviation of a factor of $\sim \! 1/3$
(\cite{Kawaguchi10}).

\section{Anisotropic Emission of Accretion Disk}

Radiation flux ($F$) from 
 a unit surface area of the disk toward a unit 
 solid angle at the polar angle of $\theta$ 
 decreases with an increasing $\theta $ as follows:
\begin{equation}
F \propto \cos \theta \, (1 + 2 \cos \theta ).
 \label{eq:flux}
\end{equation}
Therefore, an accretion disk emits lesser 
radiation in the direction closer to its equatorial plane 
(i.e., larger $\theta $; 
\cite{Laor89}, \cite{Sun89}; Figure~1, right). 
(Relativistic effects, such as light bending near the BH, 
unlikely alter this anisotropy drastically.)
Thus, the assumption of isotropic emission from accretion disks 
(e.g., eq.\ref{eq:bar87})
obviously overestimates the radiation flux toward the 
torus, and therefore overestimates the torus inner radius.

\section{Inner Structure of Dusty Torus}

\begin{figure}
\center
\includegraphics[scale=0.55,bb=66 171 745 560]{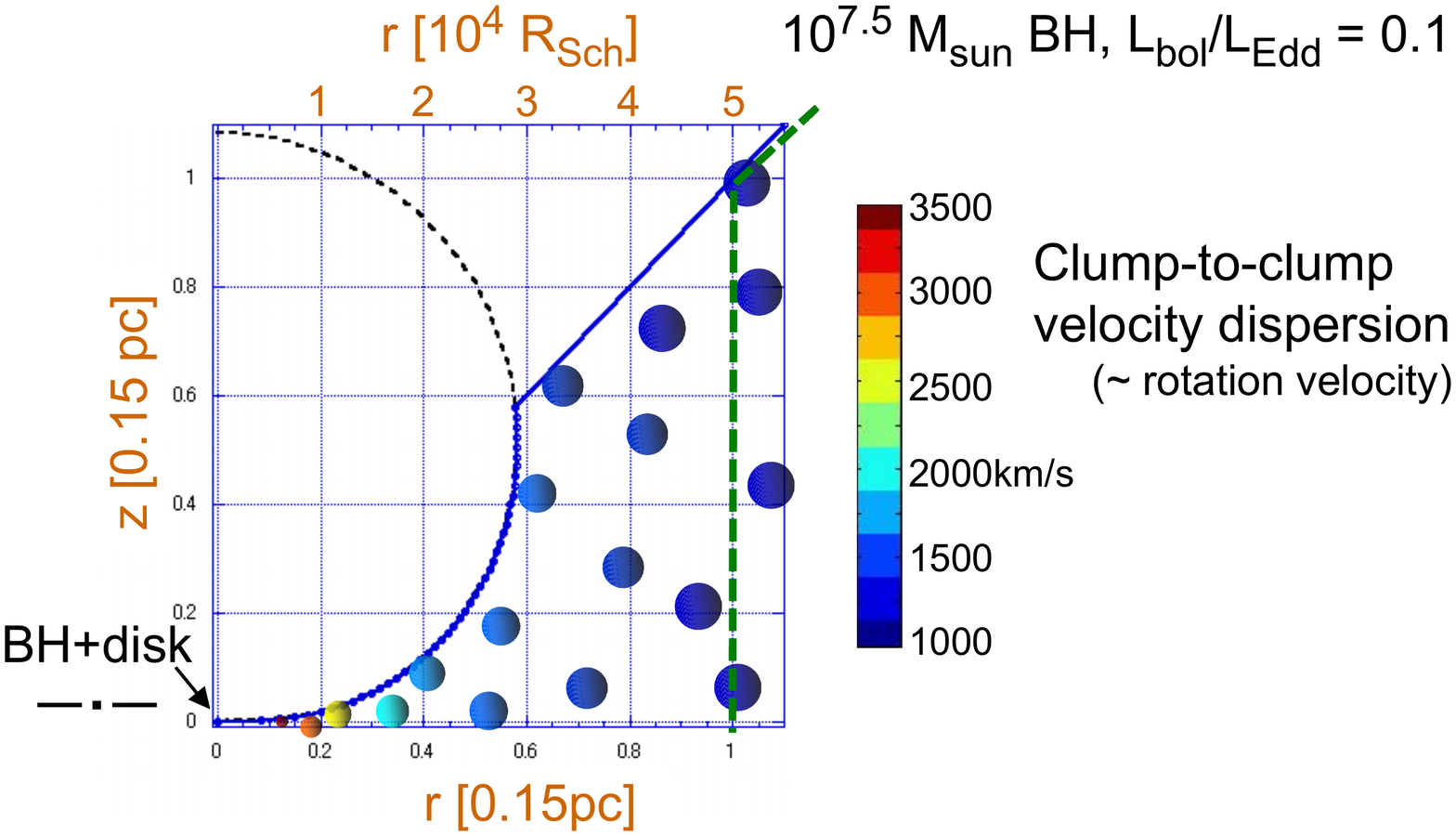}
\caption{\label{fig2} 
Inner region of the torus in the cylindrical coordinates ($r$, $z$) \cite{Kawaguchi10}.
At the left-bottom corner, let us suppose that there are 
the central BH and the accretion disk.
Thick blue solid curve indicates the edge of the torus, with 
the opening angle of the torus assumed to be 45\,deg.
On the right-hand side of the line, the temperatures of clumps 
are below the dust sublimation temperature.
If an observer assumes that the disk illumination is isotropic, 
then he/she would expect the torus inner edge at $R_{\rm sub,0}$ 
(thick, green dashed line).
Units of the scale (in pc and Schwarzshild radius $R_{\rm Sch}$) 
are computed for a $10^{7.5} M_\odot$ BH, with 
$L_{\rm bol}/L_{\rm Edd}$ of 0.1 and $L_{\rm bol}$ of $\sim 3 L_{\rm UV}$.
Different colors of clumps indicate the clump-to-clump 
velocity dispersion of the order of the rotational velocity $\sim$1000\,km/s.
Temperature of clouds are less than or equal 1500\,K, corresponding to 
$\sim 4$km/s.
The actual size of clumps are expected to be $\sim 10^{-3}$ of the 
distance to the BH, 
but drawn with much bigger sizes here for illustration purposes.
}
\end{figure}

We determine the inner edge of the torus so that 
the temperature of a clump (at the irradiated surface) equals 
to the sublimation temperature 
at the edge.
As mentioned above, radiation flux from the accretion disk varies
with the polar angle $\theta $.
Thus, the sublimation
radius of the torus is also a function of 
$\theta $, which is indicated by $R_{\rm sub}(\theta )$. 
Although various grain sizes result in the sublimation 
process occurring over a transition zone rather than a single 
distance (\cite{Nenkova08}), we employ a sharp boundary 
for simplicity.

Figure~2 
shows the calculated structure 
of the innermost region of the torus.
The central BH and the accretion disk on the $z=0$ plane are located at the 
origin of the coordinate axes (at the left-bottom corner). 
The actual size of clumps are expected to be $\sim 10^{-3}$ of the 
distance to the BH
(\cite{Vollmer04}, \cite{Honig07}, 
see Apendix of \cite{Kawaguchi11}), 
but drawn as if they are much bigger for illustration purposes in this figure.

It turns out that (i) the torus inner edge is located closer to 
the central BH than suggested by previous estimations (eq. \ref{eq:bar87})
and that (ii) the structure of the edge is concave/hollow.
Moreover, (iii) $R_{\rm sub}(\theta )$ decreases down to 
$0.1 \times R_{\rm sub,0}$ at $\theta =88.5$deg. 
This radius coincides with the outermost radius of AGN accretion 
disks, which is determined by the onset of radial  
self-gravity of the disk (see \cite{Kawaguchi04}).
Our result indicates that there is no gap between the torus 
and the disk. 

\section{Transfer Function}

The innermost radius of the torus discussed above cannot be spatially 
resolved even 
in the coming decade.
Thus, observations of time variability will continue to be 
powerful tools to explore the innermost structure of the torus. 
In this section, we calculate the 
time variation of NIR emission in response 
to a $\delta$-function like variation of the 
irradiation optical/UV flux [the transfer function $\Psi (t)$]. 

To calculate $\Psi (t)$ for the clumpy torus, we consider the following 
items;
(1) the optical path, 
(2) NIR emissivity of the torus inner region as a function of $\theta $
and (3) anisotropic emission of each clump.
(4) 
We also include the effect of the torus 
self-occultation (i.e., absorption of NIR emission from a clump 
by other clumps on the line of sight). 
While the torus self-occultation is 
a minor effect for a typical type-1 AGN, 
it plays a significant role 
for 
inclined viewing angles, thick tori 
and misaligned tori.

Our calculation results explain the observed propoerties 
of the existing data (e.g., lag-$L$ relations), and 
present theoretical predictions that will be tested 
observationally in the future.
Details are given in \cite{Kawaguchi11}.
Below, we show some examples of our calculation results shortly.

\subsection{Viewing angle dependency}

\begin{figure}
\center
\includegraphics[scale=0.65,bb=71 228 774 563]{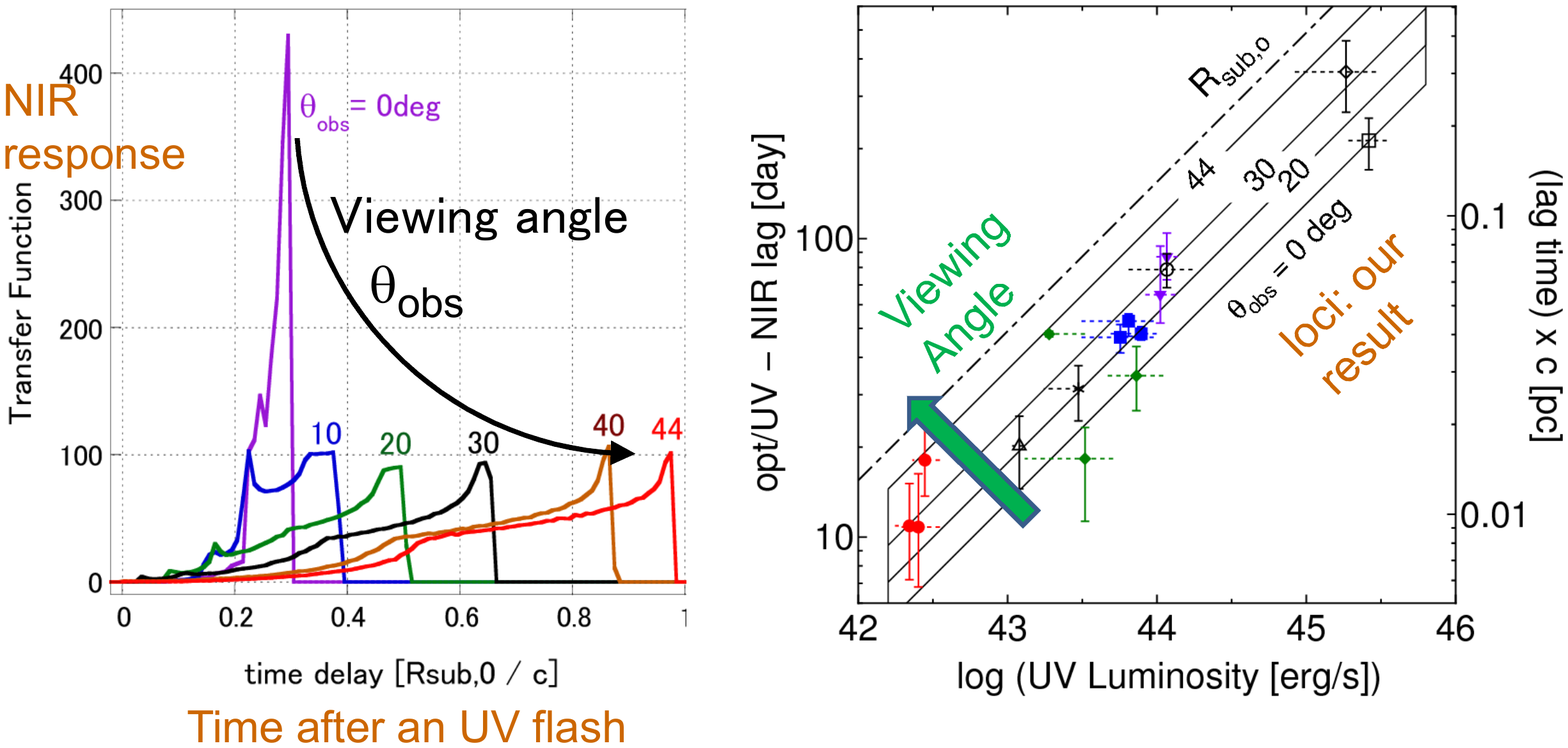}
\caption{\label{fig3} 
Left: Transfer functions for various viewing angles $\theta_{\rm obs}$, 
from a face-on view (purple) to intermediate type-1 cases (orange and red) \cite{Kawaguchi11}.
The time delay appears longer for more inclined angles.
Right: Our results (loci) on the time delay v.s.\ UV luminosity plane.
Observed points are taken from \cite{Suganuma06}, and 
covered well by our calculational results.
}
\end{figure}

Figure~3 (left panel) 
shows the transfer functions for various viewing angles 
$\theta_{\rm obs}$, from an exactly pole-on geometry 
($\theta_{\rm obs}=0$\,deg) to 
inclined viewing angles.
With a large $\theta_{\rm obs}$ 
($\approx 40$--$44$deg), 
the line of sight grazes the upper boundary of the torus, 
which would corresponds to the situation in type-1.8/1.9 AGNs.

In order to compare our results with the observed data 
more directly, 
the $t_{\rm delay}$ v.s. $L_{\rm UV}$ diagram 
is drawn in Figure 3 (right panel).
Our loci for various $\theta_{\rm obs}$ 
well cover the observed scatter.
Namely, our model expalins not only the systematic shift 
(by a factor of 3) on average, but also the scatter aroud 
the regression line.

\subsection{Accretion rate dependency}

\begin{figure}
\center
\includegraphics[scale=0.55,bb=80 236 760 561]{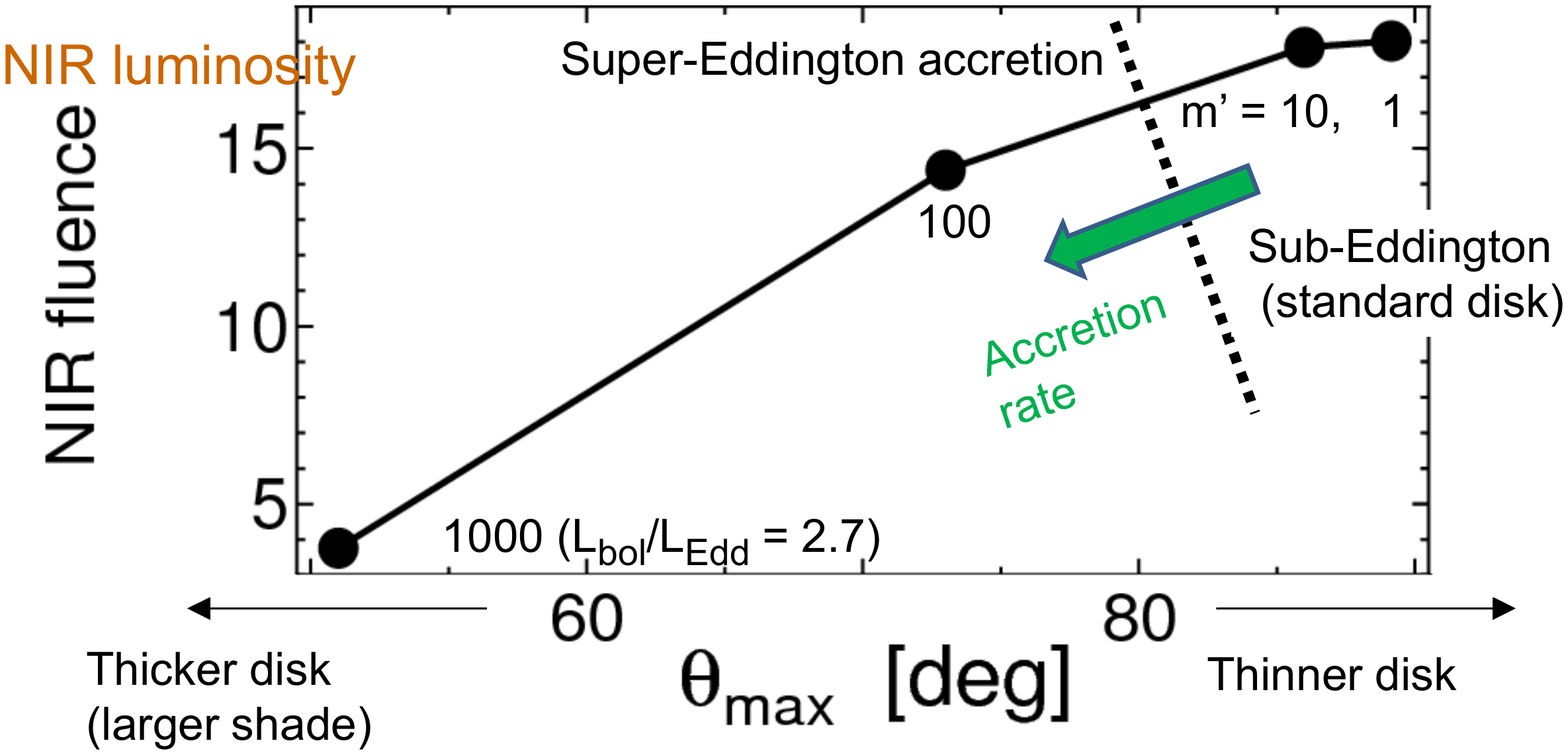}
\caption{\label{fig4} 
NIR fluence (an integration of the transfer function over time)
as a function of the disk thickness (with $\theta_{\rm max}$ 
being the angle at the disk surface measured from the rotational 
axis towards the midplane at 90\,deg) \cite{Kawaguchi11}.
As the accretion rate $\dot{M}$ (where $\dot{m}$ represents 
$\dot{M}$ in the unit of $L_{\rm Edd}/c^2$, with 
$\dot{m}$ of 16 producing the Eddington luminosity $L_{\rm Edd}$), 
the disk shade becomes large.
Then, the expected NIR flux for the cases with 
high accretion rates (e.g.,  $\dot{m} = 1000$, shining at 
$2.7 L_{\rm Edd}/c^2$, see \cite{Kawaguchi03}) reduces 
down to $\sim 1/5$ of the NIR flux with sub-Eddington accretion rates.
}
\end{figure}

When the accretion rate exceeds the Eddington rate
($\approx \! 16 L_{\rm Edd}/c^2$), 
an optically thick
advection-dominated accretion flow (a slim disk)
appears (\cite{Abramowicz88}). 
Since super-Eddington disks are geometrically thick 
(\cite{Abramowicz88}), 
they cannot illuminate the directions near their 
equatorial plane by the disk self-occultation (\cite{Fukue00}).
Moreover, the observed data do not support the concept of 
Eddington-limited accretion (\cite{Collin04}).
Thus, a strong anisotropy of the disk emission, 
such as Equation (\ref{eq:flux}) and the disk self-occultation, 
is required to allow gas infall to super-Eddington 
accreting sources.

As discussed in \cite{Kawaguchi10}, some AGNs 
with presumably super-Eddington accretion rates 
show the weak NIR emission  
(Ark564, TonS180, J0005 and J0303; 
\cite{Rodriguez06}, 
\cite{Kawaguchi04}, 
\cite{Jiang10}; see, however, \cite{Hao10}).
Observationally, it is controversial whether high Eddington ratios 
are related to the weak NIR emission or not.
Therefore, we here briefly summarize what are the expected 
behaviours from the theoretical sides (\cite{Kawaguchi11}).

We deduce the disk thickness at the Far-UV emitting region, 
based on the work 
by \cite{Kawaguchi03}.
Although more detailed calculations of the accretion disk 
are possible (e.g., magneto-hydrodynamical simulations rather 
than the $\alpha$-description for the viscosity as used in 
\cite{Kawaguchi03}),
it is a natural consequence that the disk puffs up as the 
accretion rate increases irrespective of calculational methods.
We do not think that such time consuming calculations 
of the disk have scientific gains here.

Larger accretion rates make the disk thicker and the 
shade of the disk itself larger. 
Thus, the NIR fluence becomes small as the accretion rate increases. 
This is consistent with the weakness of the 
X-ray emission line from neutral iron, which likely originates 
in the illuminated torus, 
of objects with high Eddington ratios (\cite{Bianchi07}).

Super-Eddington accretion rates cause another influence upon 
the NIR flux.
When the accretion rate becomes super-Eddington, 
the disk self-gravity starts to govern the disk and 
truncate the outer part of the disk.
Due to the truncation, 
super-Eddington disks 
do not radiate at NIR (\cite{Kawaguchi04}).

In contrast to the result for a thin disk with a sub-Eddington
accretion rate, 
a super-Eddington accretion rate leads to a much weaker 
NIR emission because of the two effects above. 
In other words, object selections with rest-NIR emission tend 
to miss super-Eddington accretors.

\subsection{Summary and prospects}

We have investigated the consequences of the anisotropic nature 
of the disk emission (\cite{Kawaguchi10}, \cite{Kawaguchi11}).
We found that the 
torus inner region is much close to the disk and concave.
Also, the anisotropy natually resolves the conflict in the 
NIR time delay -- luminosity relation between theory and observations.
Larger inclination angles (corresponding to intermediate type-1 
AGNs, such as type 1.9) lead to longer time delays.
Large accretion rates result in the weak NIR emission, due to 
the large disk shade and the disk truncation.
Therefore, we tend to miss super-Eddington accretors 
when we include the NIR flux to select objects from survey data. 
More predictions for other quantities (such as variability amplitudes) 
and for other dependencies (torus thickness and torus-disk misalignment) 
are presented in \cite{Kawaguchi11}.

Now, we mention the prospects for future X-ray mission Astro-H, 
which will have a high spectral resolution of the order of $100$km/s.
Figure~2 shows the expected velocity of clumps in the 
torus by various colors.
At the inner edge of the torus, clumps will have velocity 
dispersion of $\sim 1500-3000$km/s. 
Thus, such clump kinematics will be unveiled by Astro-H.

%
%
\small  
%
%

%

%
\end{document}